\documentclass[pre,aps,twocolumn,floatfix,superscriptaddress,showpacs]{revtex4}

\usepackage{graphicx}
\usepackage{dcolumn}
\usepackage{amssymb}
\usepackage{amsmath}
\usepackage{latexsym}
\usepackage{verbatim}
\usepackage{color} 

\begin{document}

\title {
  Scaling and entropy in $p$-median facility location along a line
}

\author{Michael T.~Gastner}
\affiliation{Department of Mathematics, Complexity and Networks Programme,
  Imperial College London, South Kensington Campus, London SW7 2AZ, 
  United Kingdom}

\begin{abstract} 
  The $p$-median problem is a common model for optimal facility location. 
  The task is to place $p$ facilities (e.g., warehouses or schools)
  in a heterogeneously populated space such that the average distance from a 
  person's home to the nearest facility is minimized.
  Here we study the special case where the population lives along a line 
  (e.g., a road or a river).
  If facilities are optimally placed, the length of the line segment served 
  by a facility is inversely proportional to the square root of the 
  population density.
  This scaling law is derived analytically and confirmed for concrete 
  numerical examples of three US Interstate highways and the Mississippi
  River.
  If facility locations are permitted to deviate from the optimum, the number 
  of possible solutions increases dramatically.
  Using Monte Carlo simulations, we compute how scaling is affected by an
  increase in the average distance to the nearest facility.
  We find that the scaling exponents change and are most sensitive near the 
  optimum facility distribution.
\end{abstract}
\maketitle

Quantitative studies in many branches of science frequently reveal scaling
laws where two sets of observables are related by a power law over several
orders of magnitude.
Examples range from astronomy (e.g., Kepler's third law) to biology where,
for example, Kleiber's law states that the metabolic rates of mammals scale 
approximately as the three-quarter power of their body mass~\cite{Kleiber32}.
Here we look at a problem from economic geography, the relationship between the
spatial distribution of a population and the distribution of service
establishments (e.g., post offices or gas stations). 

Physicists typically enjoy the luxury of measuring scaling exponents in
carefully designed and repeatable experiments.
In biology and the social sciences, by contrast, the exact circumstances of
an experiment are generally more difficult to control and to repeat.
As a consequence, power-law exponents are frequently obfuscated by noise in 
the measurement and in the process generating the scaling law itself.
The remaining uncertainty can lead to heated debates if, for example, the
scaling exponent in Kleiber's law is not truly $2/3$ instead of 
$3/4$~\cite{Dodds_etal01,WhiteSeymour03}.
The available geographic data for the distribution of service establishments
leave similar room for interpretation so that various scaling laws have been
proposed~\cite{Stephan77,Stephan88,Bettencourt_etal07,Um_etal09}.

Facing such controversies, theorists often try to calculate the ``correct''
exponent from deterministic models.
One recurring idea is that scaling should emerge naturally from some 
appropriate model if an objective function (energy 
dissipation~\cite{West_etal97}, earnings~\cite{GabaixLandier08}, 
travel distance~\cite{Stephan77,Gusein93,GastnerNewman06}, etc.) is 
optimized. 
This approach has led to elegant theories, but it leaves one key problem 
unaddressed.
Knowing that evolutionary biology, human decisions, or other processes
shaping the available empirical data are intrinsically stochastic, 
there is in principle a huge variety of outcomes.
How many different solutions are conceivable? 
How close to optimal does the observed solution need to be in order to exhibit 
the theoretically predicted scaling exponent?

Here we study a model which serves as an example of computational techniques 
suited to address these questions.
The model is the $p$-median problem of optimal facility location along a
strongly heterogeneously populated line (e.g., a transcontinental highway).
The task is to place $p$ facilities along the line and find the configuration 
that minimizes an objective function, in this case the average distance to 
the nearest facility~\cite{HassinTamir91}.
Ignoring small-scale heterogeneity in the population, an analytic calculation 
predicts a simple scaling law for the length of the line segments served by 
different facilities.
The exact optimum locations can be computed numerically for realistic input 
data and are in good agreement with the analytic prediction.
Using techniques from statistical physics, we calculate the number of possible 
facility locations for non-minimal costs.
With Monte Carlo simulations we will then quantify how deviations from the 
optimum make it less likely to find the theoretical exponent.

\section{The $p$-median problem}

The challenge in facility location problems is to place $p$ service centers 
or facilities so that $n$ demand points are optimally served (see for example
Ref.~\onlinecite{DreznerHamacher02} for an overview).
Facilities can be hospitals, supermarkets, fire stations, libraries, 
warehouses, or any other supply centers providing vital resources to the 
population living at the demand points (e.g., households or cities).
Here we consider the case where the demand points are at regular intervals 
along a one-dimensional geographic object, such as a road or a river, and
where every demand point is a possible location for a facility.
The number of people $N$ who require the facilities' services is assumed to 
be known at each demand point.
This number is typically very heterogeneous across geographic space.
Depending on the context, there are different strategies for the placement 
of the facilities.
In this article, we concentrate on the $p$-median problem, an important
special case, where the objective is to minimize the average distance
between a person's demand point and the nearest facility.
(A recent summary of the vast literature on the $p$-median problem can be
found in Ref.~\onlinecite{Reese06}).

\begin{figure}
  \begin{center}
    \includegraphics[width=8.6cm]{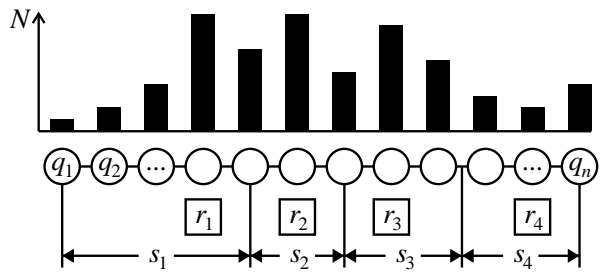}
    \caption{Illustrative example of the $p$-median problem along a line.
    The population $N$ is known at the demand points $q_1, \ldots, q_n$.
    In this article, the distance between neighboring demand points is 
    assumed to be constant.
    Facilities will be placed on $p$ of these $n$ demand points.
    (In the figure, $p=4$.)
    Their locations $r_1, \ldots, r_p$ are to be determined so that the 
    average distance between a demand point and the nearest facility,
    weighted by $N$, is minimized.
    After the facilities have been located, the line can be divided into $p$ 
    segments $s_1, \ldots, s_p$ so that the $i$-th segment corresponds to the 
    service region of the $i$-th facility.
    }
    \label{pmed_intro}
  \end{center}
\end{figure}

Let us call the facility locations from left to right $r_1, \ldots, r_p$.
These positions are chosen among the demand points $q_1, \ldots, q_n$, which 
are equidistant (i.e., $q_{i+1}-q_i = \text{const.}$ for $i=1, \ldots, n-1$) 
along a line (see Fig.~\ref{pmed_intro}).
If the population at $q_i$ is denoted by $N_i$, the $p$-median problem consists
of minimizing the cost function~\footnote{
Because $N_i$ for a given location problem is constant, we could in principle
directly minimize the numerator in Eq.~\ref{pmedian} and ignore the 
denominator. We have decided to keep the denominator so that $C$ equals the 
average distance. $C$ can then be more easily compared across
different location problems.
}
\begin{equation}
  C(r_1,\ldots,r_p) =
  \frac{\sum_{i=1}^{n} N_i \min_{j=1,\ldots,p}|q_i-r_j|}{\sum_{i=1}^{n} N_i}.
  \label{pmedian}
\end{equation}

Because only trips to the nearest facility play a role in Eq.~\ref{pmedian},
the line along which the demand points are located can be partitioned into
$p$ segments or service regions.
Demand points belong to the same segment if and only if they share the same 
closest facility, see Fig.~\ref{pmed_intro}.
The length of facility $i$'s service region is given by
\begin{equation}
  s_i = 
  \begin{cases}
    \frac12 (r_1 + r_2)-q_1 & \text{if $i=1$},\\
    \frac12 (r_{i+1} - r_i) & \text{if $i=2, \ldots, p-1$},\\
    \frac12 (r_{p-1} + r_p)-q_n & \text{if $i=p$}.
  \end{cases}
\end{equation}
We will now take a closer look at the relation between $s_i$ and the
population density around facility $i$.

\section{Scaling of the lengths of the service regions}

At first sight, it is plausible that the spatial density of facilities should
follow the same trend as the population density: where there are more people
there should be proportionately more facilities.
However, as we will see shortly, the $p$-median solution does not follow this
rule that would give every facility an equal number of customers.
Instead facilities are less abundant per capita in the high-demand regions 
than in the low-demand regions.

\begin{figure}
  \begin{center}
    \includegraphics[width=8.6cm]{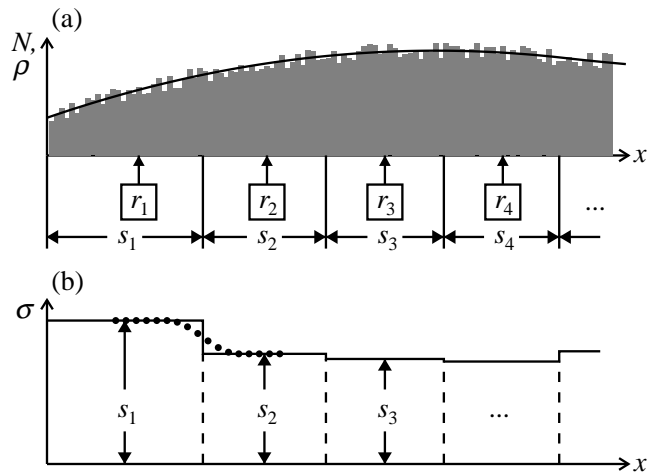}
    \caption{(a) Under the assumption that the population $N$ (gray histogram)
      varies little between neighboring demand points, $N$ can be approximated
      by a continuous function $\rho$ (black curve).
      (b) The function $\sigma(x)$ is defined as the length $s$ of the 
      segment covering position $x$.
      Strictly speaking, $\sigma$ is a piecewise constant function.
      However, if the spatial variations in $N$ are sufficiently small,
      $\sigma$ can be approximated by a continuous function
      (indicated by the dotted curve).
    }
    \label{cont_approx}
  \end{center}
\end{figure}

For a spatially heterogeneous population distribution $N_i$, it is difficult 
to deduce this general trend directly from Eq.~\ref{pmedian}.
With certain approximations, however, the problem becomes analytically
tractable; essentially, we translate the line of reasoning developed in
Ref.~\onlinecite{Gusein93} and \onlinecite{GastnerNewman06} for the 
two-dimensional $p$-median problem to the one-dimensional case.
First we define the population density $\rho(x)$ which is the number of people
per unit length in the vicinity of $x$.
Equation~\ref{pmedian} can be rewritten as
\begin{equation}
  C(r_1, \ldots, r_p) =
  \frac{\int_{q_1}^{q_n} \rho(x)\min_{j=1,\ldots,p}|x-r_j|\,dx}
  {\int_{q_1}^{q_n} \rho(x)\,dx},
  \label{pmed_cont_approx}
\end{equation}
where we have used the new notation to replace sums by integrals.
If we allow $\rho$ to be piecewise constant, this expression is still
exact, but later it will be more convenient to approximate 
$\rho$ with a continuous function (Fig.~\ref{cont_approx}a).

Next we define $\sigma(x)$ to be the length of the
segment serviced by the facility closest to $x$ (see Fig.~\ref{cont_approx}b).
The average distance from facility $j$ to a point $x$ inside its service
region is equal to $g_j \sigma(x)$, where $g_j$ depends on the exact location of
the facility.
For example, if $r_j$ is close to the center of the segment, 
$g_j\approx\frac14$.
In the spirit of a mean-field approximation, we will now assume that $\rho$
varies little over the size of a segment.
Then we can replace the exact distance, $\min|x-r_j|$, in the numerator of 
Eq.~\ref{pmed_cont_approx} with its average $g_j\sigma(x)$,
\begin{equation}
  C \approx \frac{\int_{q_1}^{q_n} \rho(x)\,g\,\sigma(x)\,dx}
  {\int_{q_1}^{q_n} \rho(x)\,dx}.
  \label{pmed_mf_approx}
\end{equation}
The index $j$ was dropped in Eq.~\ref{pmed_mf_approx} assuming that most 
facilities will be close to the center of their service region so that
$g_j$ is approximately constant.

Unlike in Eq.~\ref{pmedian}, the locations $r_j$ no longer appear explicitly 
in Eq.~\ref{pmed_mf_approx}.
Instead we have to find the function $\sigma(x)$ that minimizes $C$ subject to
the constraint that there are $p$ facilities.
This constraint can be expressed as
\begin{equation}
  \int_{q_1}^{q_n} \frac1{\sigma(x)}dx = p.
  \label{constraint}
\end{equation}
Introducing a Lagrange multiplier $\alpha$, the problem is equivalent to
finding the zero of the functional derivative
\begin{equation}
  \frac{\delta}{\delta \sigma}
  \left[\frac{g\int_{q_1}^{q_n} \rho(x)\,\sigma(x)\,dx}
    {\int_{q_1}^{q_n} \rho(x)\,dx}
    -\alpha\left(p-\int_{q_1}^{q_n}\frac1{\sigma(x)}dx\right)\right] = 0,
\end{equation}
solved by
\begin{equation}
  \sigma(x) = \sqrt\frac{\alpha\int_{q_1}^{q_n}\rho(x')\,dx'}{g\,\rho(x)}.
\end{equation}
The Lagrange multiplier can be eliminated by inserting this expression into
Eq.~\ref{constraint}.
After some algebra,
\begin{equation}
  \sigma(x) = \frac{\int_{q_1}^{q_n} \sqrt{\rho(x')} dx'}{p\sqrt{\rho(x)}}
  \propto [\rho(x)]^{-1/2}.
  \label{sqrt_scaling}
\end{equation}

The lengths of the service regions are thus inversely proportional to the
square root of the population density.
The spatial density of facilities $1/\sigma$ increases $\propto \rho^{1/2}$, 
but the per-capita density $1/(\rho\,\sigma)$ decreases $\propto \rho^{-1/2}$ 
with growing population.
The square-root scaling is a compromise providing most services where they 
are most needed, namely in the densely populated regions, but still leaving
sufficient resources in sparsely populated regions where travel distances are
longer.
This result implies an economy of scales: In crowded cities
fewer facilities per capita can supply a larger population than in rural
areas.
If facilities and demand points are not restricted to be along a line, but 
can be placed in two-dimensional space, the scaling exponent is $2/3$
instead
of $1/2$~\cite{Gusein93,GastnerNewman06} (see Section~\ref{conclusion}).
However, economies of scale are also predicted in two dimensions.
Empirical studies have indeed reported this effect for certain classes of real
facilities~\cite{Stephan77,Bettencourt_etal07,Um_etal09}.

\section{Exact solution for empirical population distributions}
\label{exact_numerical}
\begin{figure}
  \begin{center}
    \includegraphics[width=8.6cm]{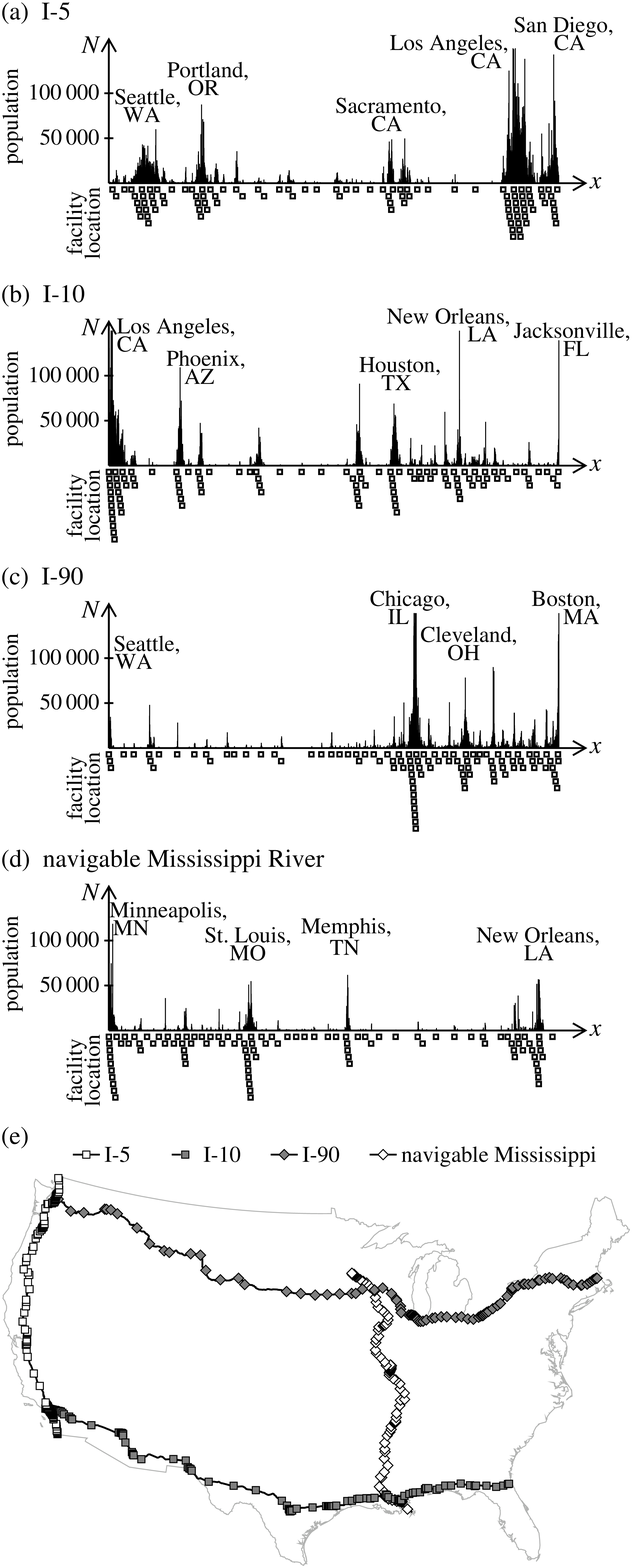}
    \caption{Population $N$ as a function of position $x$ along (a) 
      Interstate 5, (b) 10, (c) 90, (d) the navigable part of the
      Mississippi River.
      The small squares below the $x$-axes indicate the optimal $p$-median
      positions of $100$ facilities.
      (e) Map of the roads, the river, and the facility locations.
    }
    \label{test_sets}
  \end{center}
\end{figure}

The calculation in the previous section assumes that the population density 
$\rho(x)$ varies little within a service region.
As we can see from Eq.~\ref{sqrt_scaling}, this implies that the segment
length $\sigma(x)$ is also a smooth function (Fig.~\ref{cont_approx}b).
Real census data, however, typically reveal strongly varying populations
even on small spatial scales.
In Fig.~\ref{test_sets}a--d, we show population numbers near three US 
Interstate highways and the navigable Mississippi River.
The data were generated from the US census of the year 2000.
First, Interstates 5, 10, 90 and the Mississippi River were 
parameterized by arc length and markers were placed at regular 1-km intervals. 
Then census blocks within 10 km of the highways or the Mississippi were 
identified and their population assigned to the nearest kilometer marker.
As is clear from Fig.~\ref{test_sets}a--d, neither of the four populations is 
a smooth function.
Whether the assumptions behind Eq.~\ref{sqrt_scaling} are valid, is questionable,
but it turns out that the scaling law for the service regions still holds with
surprising accuracy.

\begin{figure}
  \begin{center}
    \includegraphics[width=8.6cm]{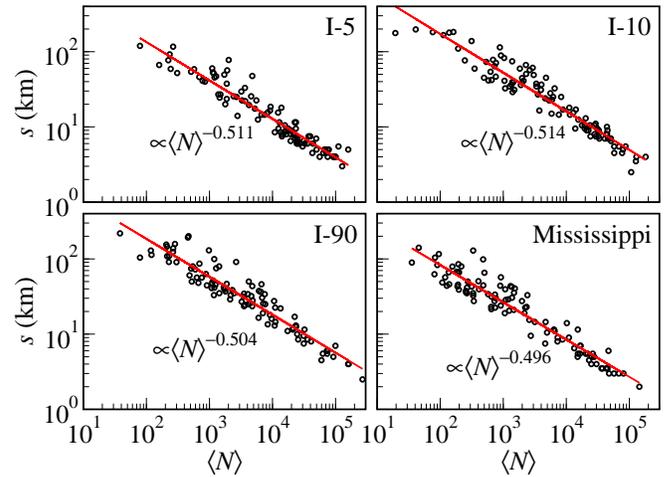}
    \caption{(Color online) The length of a service region $s$ versus
      the mean population $\langle N\rangle$ in this service region.
      Lines indicate least-squares fits to Eq.~\ref{log-log-scaling}.
      Scaling is in good agreement with the analytic prediction 
      $s\propto\langle N\rangle^{-1/2}$.}
    \label{scaling_exact_opt}
  \end{center}
\end{figure}

To compute the scaling exponent, $p=100$ facilities are placed on each of the
four test data sets.
The optimal locations are calculated with the efficient algorithm of
Ref.~\onlinecite{HassinTamir91}.
Their positions along the roads and the river in geographic space are shown 
in Fig.~\ref{test_sets}e.
The segment lengths $s_i$ are calculated for each facility $i=0,\ldots,p$.
In Fig.~\ref{scaling_exact_opt}, $s_i$ is plotted versus the mean value of 
$N$ inside the segment, denoted by $\langle N\rangle_i$
\footnote{If both the left and the right boundary, $b_l$ and $b_r$, of the 
segment are half-integers, $\langle N\rangle_i$ is defined as 
$\sum_{j=\lceil b_l+1/2\rceil}^{\lfloor b_r-1/2\rfloor}N_j/s_i$
If $b_l$ or $b_r$ is an integer, $N_{b_l}/(2s_i)$ or $N_{b_r}/(2s_i)$ is
added to the sum (i.e., half of the population is assigned to the facility on 
the right, half to the equally distant facility on the left).
}.
Ordinary least-squares fits of 
\begin{equation}
  \log(s_i) = a\log\langle N\rangle_i + \text{const.}
  \label{log-log-scaling}
\end{equation}
to the data yield slopes $a=-0.511$ (I-5), $-0.514$ (I-10), $-0.504$ (I-90), 
and $-0.496$ (Mississippi River), close to the prediction $a=-1/2$ of 
Eq.~\ref{sqrt_scaling}.
The correlations are strong; $R^2$ is consistently bigger than $0.89$.
Assuming that the residuals are log-normally distributed, the predicted
value $-1/2$ is in all cases within the 95\% confidence intervals.
Thus, the equivalent of Eq.~\ref{sqrt_scaling}, 
$s_i \propto (\langle N\rangle_i)^{-1/2}$,
obtained by replacing the continuous variables $\rho$ and $\sigma$ by their
discrete counterparts $\langle N\rangle_i$ and $s_i$, is a good approximation.
This observation demonstrates that scaling at the exact $p$-median 
configuration is robust
even in the presence of strong spatial fluctuations.

\section{The number of configurations for non-minimal costs}
That the square-root scaling of the service regions is discernible even
for realistically heterogeneous input, establishes a potential link to
previous empirical work.
Data collected in Ref.~\onlinecite{Stephan88,Bettencourt_etal07,Um_etal09} 
suggest, at least for certain classes of facilities, a sublinear dependence 
of service facilities on population numbers.
It has been conjectured that the $p$-median model~\cite{Stephan77} or a 
generalization thereof~\cite{Stephan88,Um_etal09} might explain this trend.
Admittedly, we are looking in this article at a simplified linear geometry.
Yet that sublinear scaling is robust even for substantially noisy input, 
might be viewed as supporting evidence for this conjecture.

However, there is more to the problem than first meets the eye.
Although it is mathematically convenient to assume that facilities are
placed to minimize an objective function such as Eq.~\ref{pmedian}, it is far 
from clear that the exact minimum will be achieved in reality.
Decisions about facility locations are probably more haphazard in real life.
For example, site selections may be swayed by political interests, short-term 
fluctuations in property prices, or based on an incomplete knowledge of
the actual demand.
Even if the best effort is made to reach the global optimum, ``accidents of 
history'' may keep the facility locations trapped in a costlier local optimum.
It seems overly optimistic to draw conclusions about the scaling of real 
service regions only from the best of all solutions.
The available literature for real facility 
distributions~\cite{Stephan88,Bettencourt_etal07,Um_etal09} -- rather than
the numerically optimal ones discussed in Sec.~\ref{exact_numerical} --
also justifies cautious skepticism, as some significant differences to the
$p$-median result have been observed in reality, albeit in two dimensions.

How many facility configurations with costs near, but not necessarily equal to,
the global minimum exist?
There is no simple way to answer this question.
Although the algorithm of Ref.~\onlinecite{HassinTamir91} can find the global 
optimum very efficiently, it does not provide information about non-optimal 
solutions.
Scanning all possible configurations is out of the question because their 
number is too vast.
Even for our smallest test data set (I-5) there are 
$\binom{2213}{100}\approx 3.5\cdot 10^{175}$ different ways to locate the 
facilities.
The situation is reminiscent of many-particle systems in physics where one 
wishes to calculate the large number of micro-states at a certain energy 
level out of an even larger number of all conceivable micro-states.
In that context, statistical mechanics has developed many powerful numerical
tools.
We will build on this analogy in order to estimate the number of 
non-optimal facility locations.

Let us call $\Omega(C)dC$ the number of facility locations with costs between 
$C$ and $C+dC$.
The function $\Omega(C)$ plays the role of the ``density of states'' in 
statistical mechanics.
As we will see, $\Omega$ increases very rapidly as $C$ exceeds the minimum
$C_{min}$, so that it will be more convenient to work with its logarithm,
the entropy $S(C)=\log\Omega(C)$.
Our aim is to calculate $S$ with Monte Carlo simulations.
Several methods exist~\cite{Lee93,deOliveira_etal96,Argollo_Lima03};
here we apply the Wang-Landau algorithm~\cite{WangLandau01a}.
First, the range of possible costs is divided into small discrete intervals of 
length $\Delta C$.
Then a random walk through the set of facility locations is performed and
we count, in the form of a histogram, how often each interval is visited.
The main idea behind the Wang-Landau algorithm is to bias the random walk in
such a manner that all intervals are visited equally often.
For such a ``flat histogram'' we obtain equally good statistics for all 
intervals, an advantage when $S(C)$ is the basis of further calculations.
We describe details of our implementation in App.~\ref{wang_landau}.

\begin{figure}
  \begin{center}
    \includegraphics[width=8.6cm]{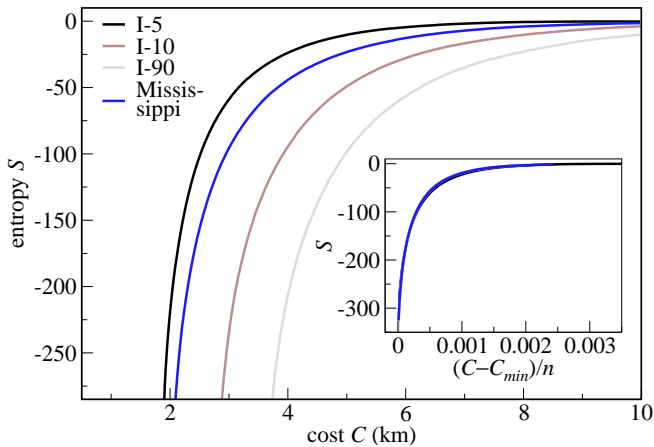}
    \caption{(Color online) The entropy $S$ (i.e., the logarithm of the density of states) 
           versus the cost $C$.
           The inset shows the same four curves as the main panel, but with
           rescaled abscissa $(C-C_{min})/n$ where $n$ is the number of
           demand points.}
    \label{entropy}
  \end{center}
\end{figure}

From calculations for four different empirical population distributions
(Fig.~\ref{entropy}) it is clear that $S$ is singular at $C_{min}$, the 
smallest possible cost.
Thus, $S$ increases enormously in the vicinity of $C_{min}$ and the density
of states $\Omega=\exp(S)$ grows even more rapidly.
The results for four different empirical population distributions suggest that
$S$ follows approximately the same curve (inset of Fig.~\ref{entropy}) if 
regarded as a function of $(C-C_{min})/n$, where $n$ is the total number of 
demand points $q_1,\ldots,q_n$.
Therefore, it appears to be a universal feature that for all realistic 
populations a large number of different possible configurations must be 
considered if the assumption of optimality is relaxed.
This observation raises the question: Can the scaling relation of
Eq.~\ref{sqrt_scaling} still be observed if facility locations are not
exactly optimal, but are among the numerous configurations achieving
almost but not exactly $C_{min}$?

\section{Is scaling detectable for non-minimal costs?}

If we randomly select a facility configuration with a cost in the interval 
$[C,C+dC]$, we can formally obtain the scaling exponent $a$ from 
Eq.~\ref{sqrt_scaling} as follows.
First, we log-transform the segment lengths $s_i$ and the population density 
$\langle N\rangle_i$.
Then a least-squares linear fit to Eq.~\ref{log-log-scaling} will be performed 
to calculate $a$.
This procedure can be coupled with the Wang-Landau algorithm so that, at
every step in the random walk through configuration space, we compute $a$,
the cost $C$, and at the end of the algorithm the mean value 
$\langle a\rangle$ as a function of $C$.

\begin{figure}
  \begin{center}
    \includegraphics[width=8.6cm]{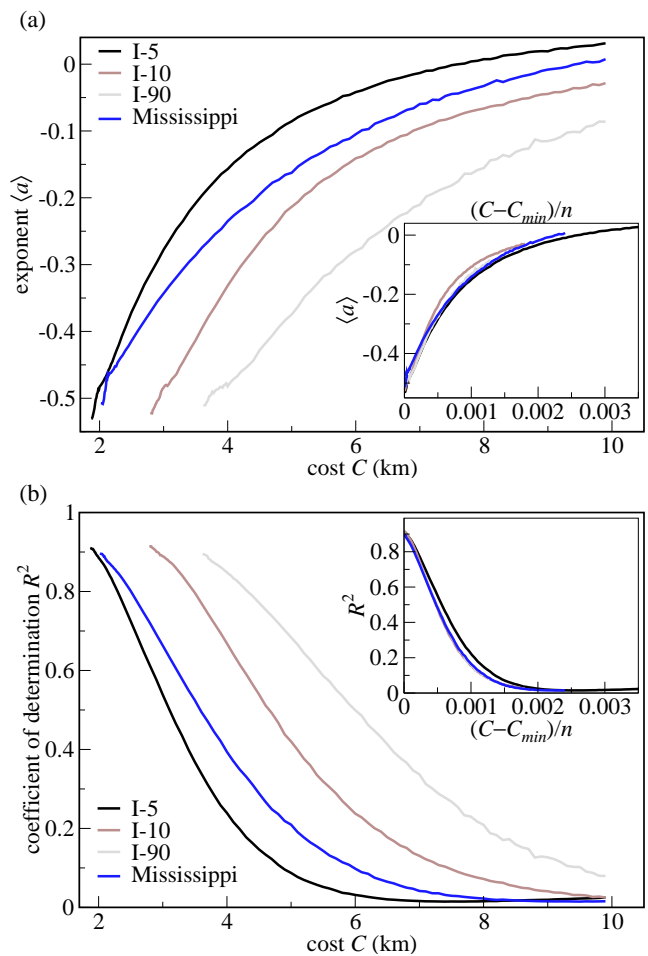}
    \caption{(Color online) (a) The mean scaling exponent $\langle a\rangle$, (b) the 
             coefficient of determination $R^2$ as a function of the cost $C$.}
    \label{exp_rsq}
  \end{center}
\end{figure}

The results, shown in Fig.~\ref{exp_rsq}a, indicate that $\langle a\rangle$ is 
approximately $-1/2$ at the minimum cost $C_{min}$ for all four numerical 
test sets, as anticipated by our earlier calculations.
As the cost increases, $a$ also increases, indicating a decreasing dependence
of the segment lengths on the population.
This behavior makes sense because the facility locations become more random 
as we move away from the optimum.
Interestingly, the overall trend how $\langle a\rangle$ increases with $C$
is similar in all four cases.
In particular, the behavior near the minimum is noteworthy because 
$\langle a\rangle$ increases most rapidly near $C_{min}$.
In other words, the analytic prediction at $C_{min}$ -- which provides us 
with the only easily calculable reference point for $a$ -- is unfortunately 
at the point where small deviations can also cause the greatest changes in 
$a$.

Together with the least-squares exponent $a$, we can also obtain other
statistical measures from linear regression, such as the coefficient of
determination $R^2$ (Fig.~\ref{exp_rsq}b).
It can take values between $0$ and $1$; the higher its value, the stronger
the correlation between $s$ and $\langle N\rangle$.
In our numerical test sets, $R^2$ takes its highest value ($\approx 0.9$) at 
$C_{min}$ and decreases as we move toward higher costs following a slightly
sigmoidal curve toward values around zero.
At very large costs, $R^2$ increases again because the solution is 
effectively an ``obnoxious facility'' location where facilities are in
sparsely populated regions so that $s$ and $\langle N\rangle$ are positively 
(instead of negatively) correlated.
For costs near $C_{min}$, however, an increase in $\langle a\rangle$ is 
coupled with a reduction in $R^2$.

\section{Conclusion}
\label{conclusion}

In this article, we have studied the one-dimensional $p$-median problem.
In one dimension, the exact optimum can be calculated numerically in 
polynomial time; in two dimensions~\cite{MegiddoSupowit84} and on arbitrary 
graphs~\cite{KarivHakimi79}, the $p$-median problem is NP-complete so that no 
polynomial-time algorithm is currently known.
Previous empirical studies of scaling in real facility locations have usually 
dealt with two-dimensional densities.
The approximate analytic result in one dimension, $\sigma\propto\rho^{-1/2}$ 
(Eq.~\ref{sqrt_scaling}), can be easily generalized for arbitrary
dimension $d$, where the size of a $d$-dimensional Voronoi cell $\sigma$ is 
predicted to scale as $\rho^{-d/(d+1)}$.
The scaling of the facility density $1/\sigma$ with the population density
$\rho$ thus remains sublinear in all dimensions.
Numerical optimization in two dimensions, based on US census data, yields 
indeed an exponent in excellent agreement with the predicted exponent 
$a=-2/3$~\cite{GastnerNewman06}.

In 1977, Stephan implicitly proposed that the $p$-median model might explain 
empirical scaling relations between the area and population density of 
subnational administrative units (e.g., states, provinces, 
counties)~\cite{Stephan77}.
Although he later generalized the objective function as more data became
available~\cite{Stephan88}, the notion that facilities may self-organize 
towards sublinear scaling has remained attractive, as proved by the recent 
rediscovery of Stephan's model by Um et~al.~\cite{Um_etal09}.

However, as the work shown here underlines, one has to be careful when 
interpreting empirical data.
Increased spatial noise in the facility distribution can lead to different
exponents and reduced correlations.
The situation investigated here portrays only one special scenario how 
randomness might be present, namely as a uniform probability distribution
over all costs in an interval $[C,C+dC]$.
It is also conceivable that not all configurations within this range are 
equally likely, so that the best-fit exponents may behave differently.
We may also replace the $p$-median model by a different optimization principle
(e.g., competitive facility location such as the Hotelling 
model~\cite{Hotelling29}) which can change the exponent at the optimum.
However, we believe that a steep increase in the number of possible 
configurations is a generic tendency of most models that relax the constraint 
of strict optimization even to a small degree.

\section{Acknowledgments}
The author thanks M.~E. Moses, S. Banerjee, B. Blasius, and H. Youn for 
stimulating discussions. The author acknowledges support from Imperial College.

\appendix
\section{Wang-Landau algorithm to calculate the density of states}
\label{wang_landau}

The Wang-Landau algorithm~\cite{WangLandau01a,WangLandau01b} is designed to 
calculate the density of states $\Omega(C)$ for $C$ in some interval 
$[C_1,C_2]$.
First the interval is divided into small sub-intervals of length $\Delta C$.
The key element of the Wang-Landau algorithm is to visit every sub-interval 
$[C,C+\Delta C]$ with a probability $\propto[\Omega(C)]^{-1}$.
Initially, the density of states is of course unknown -- this is why we need
the algorithm in the first place -- but we will recursively obtain better 
estimates for $\Omega$ as the calculation proceeds.
At the beginning we set $\Omega(C)=1$ for all intervals $[C,C+\Delta C]$.
Simultaneously we maintain a histogram $H(C)$, which counts how often a 
cost between $C$ and $C+\Delta C$ is encountered during the course of a 
random walk.
At the beginning $H(C)=0$ for all $C$.

The random walk through the set of facility locations proceeds as follows.
Starting from an arbitrary initial configuration 
$\mathbf{r}=(r_1,\ldots,r_p)$, a new set of facility positions 
$\mathbf{r'}=(r'_1,\ldots,r'_p)$ is generated with probability 
$P(\mathbf{r}\rightarrow\mathbf{r'})$.
In addition, a uniform random number $p \in [0,1]$ is generated.
If $p < \min(1,\Omega(C)/\Omega(C'))$,  the current value of $\Omega(C')$ is 
multiplied by a constant factor $f$, $H(C)$ is incremented by 1, and 
$\mathbf{r'}$ becomes the next step in the random walk.
Otherwise, the move is rejected, and we increment $\Omega(C)$ and $H(C)$
instead of $\Omega(C')$ and $H(C')$.
Following Wang's and Landau's original paper, Ref.~\onlinecite{WangLandau01a}, 
we initially set $f$ equal to the Euler number $e=2.71828..$.
When the histogram $H(C)$ is sufficiently ``flat'', $f$ is replaced by
its square root (i.e., $f\gets\sqrt f$).
For practical purposes, the histogram is treated as flat if the maximum 
number of visits recorded by $H(C)$ is less than $10\%$ more than the minimum.
If this condition is satisfied, all $H(C)$ are reset to 0, and the procedure
is iterated until $f<\exp(10^{-5}).$

From an intermediate set of facility locations $\mathbf{r}$, we generate the
new set $\mathbf{r'}$ by shifting one random facility one step to the left 
or to the right with equal probability.
Exceptions are made if the facility is already at one of the edges of the
line or adjacent to another facility.
Let us define $\nu$ to be the number of facilities on the edges ($q_1$
and $q_n$) plus twice the number of facility pairs occupying
neighboring demand points.
Then the non-zero step probabilities are given by
\begin{widetext}
  \begin{align}
    &P[(r_1, r_2, \ldots, r_p) \rightarrow (r_1-1, r_2, \ldots, r_p)] = 
      1/(2p-\nu)\;\text{if $r_1\neq q_1$},\\
    &P[(r_1, \ldots, r_{i-1}, r_i, r_{i+1}, \ldots, r_p) \rightarrow 
      (r_1, \ldots, r_{i-1}, r_i-1, r_{i+1}, \ldots, r_p)] =
    1/(2p-\nu)\;\text{if $r_i-1\neq r_{i-1}$, $i=2,\ldots,p$},\\
    &P[(r_1, \ldots, r_{i-1}, r_i, r_{i+1}, \ldots, r_p) \rightarrow 
      (r_1, \ldots, r_{i-1}, r_i+1, r_{i+1}, \ldots, r_p)] =
    1/(2p-\nu)\;\text{if $r_i+1\neq r_{i+1}$, $i=1,\ldots,p-1$},\\
    &P[(r_1, \ldots, r_{p-1}, r_p) \rightarrow (r_1, \ldots, r_{p-1}, r_p+1)] 
      = 1/(2p-\nu)\;\text{if $r_p\neq q_n$}.
  \end{align}
\end{widetext}
This set of moves is ergodic and satisfies detailed balance.

In principle, we are able to explore all costs between the globally minimal
$C_{min}$ and maximal $C_{max}$.
In practice, we have to reduce the search interval.
On one hand, $C_{max}$ is orders of magnitude larger than $C_{min}$ and we 
are interested only in $\Omega$ near $C_{min}$.
On the other hand, $\Omega$ increases so quickly that, close to $C_{min}$,
the random walk is extremely unlikely to propose a step decreasing the cost.
Therefore, we confine the random walk to intervals $[C_1,C_2]$ which become 
smaller as $C_1$ approaches $C_{min}$.
We interpolate between all estimates of $\Omega$, which all differ from
the real $\Omega$ by a multiplicative constant, with a straightforward
least-squares algorithm to obtain a single curve for the entropy 
$S=\log(\Omega)$ over all measured values of $C$.
There is exactly one constant left to be fixed because the Wang-Landau 
algorithm can calculate the entropy only up to an additive constant.
We adopt the normalization that $S=0$ at the extrapolated maximum.

\end{document}